 \documentclass[12pt,preprint]{aastex}

\shorttitle{Black Hole Mass and Eddington Ratio in QSOs}
\shortauthors{Boroson}

\begin{document}


\title{Black Hole Mass and Eddington Ratio as Drivers for the Observable 
    Properties of Radio-Loud and Radio-Quiet QSOs}


\author{Todd A. Boroson}
\affil{National Optical Astronomy Observatory, P.O. Box 26732, Tucson, AZ 85726-6732}
\email{tboroson@noao.edu}



\begin{abstract}
Recent studies of black holes in the nuclei of both active and normal
galaxies have yielded relationships that permit a physical
interpretation of the principal components of the spectra of QSOs. 
It is shown that principal component (or eigenvector) 1 (PC1) is driven
predominantly by $L/L_{Edd}$, and principal component 2 (PC2) is driven by
accretion rate.  This results in a PC2 vs. PC1 diagram in
which lines of constant black hole mass are diagonal.  Using a sample
consisting of the low-redshift PG objects supplemented by 75 radio-loud
QSOs, it is shown that such a diagram effectively distinguishes
radio-loud from radio-quiet objects as well as demonstrating that both
narrow-line Seyfert 1s and broad absorption-line QSOs lie at the high
$L/L_{Edd}$ extreme, though these two types of objects are well
separated in the PC2 direction.  A simple picture that ties together physical 
parameters (black hole mass and Eddington ratio) and classification of AGN is presented.  
Based on the location of core-dominated and lobe-dominated radio-loud QSOs, 
orientation can be modeled as a third parameter in this scheme, implying an enhancement
in the radio flux of core-dominated objects.
\end{abstract}


\keywords{galaxies: Seyfert---galaxies: nuclei---quasars: general}


\section{Introduction}

Recent studies have derived black hole masses for normal and active
galactic nuclei and have attempted to relate them to observable
properties.  Perhaps the most striking of these relations is the discovery that
the nuclear black hole mass in
normal galaxies is tightly correlated with $\sigma^n$, where $\sigma$ is the
velocity dispersion of the bulge of the host galaxy and $n$ is in the range 3.75
to 4.8 \citep{Gebhardt00, Ferrarese00}.  Although the
distances to even the nearest QSOs preclude the
spatially resolved spectroscopy that is the basis for determination of black
hole masses for normal galaxies, other techniques (e.g., reverberation
mapping), as well as the presumed extension of the relation between black hole
mass and host galaxy bulges, permit the comparison of physical parameters with
observable properties for active galactic nuclei.

\citet[BG92]{BG92} showed that most of the variance in the
measured optical emission-line properties and a broad range of continuum
properties (radio through x-ray) in a complete sample of low-redshift
QSOs was contained in two sets of correlations, eigenvectors of the
correlation matrix.  Principal component 1 (PC1) links the strength of Fe II
emission, [O III] emission, and $H\beta$ line asymmetry.
Principal component 2 (PC2) projects most strongly on optical luminosity and the
strength of He II $\lambda4686$ emission.  Subsequent studies (see \citet{Sulentic00}
and references therein) have added
observed properties to the list, and have, in general, confirmed the
reality of the correlations.

BG92 and others since have tried to understand the relationship between the 
principal components and the physical parameters that govern the energy-producing
and radiation-emitting processes.  As the summary presentation in a conference titled: 
"Structure and Kinematics of Quasar Broad Line Regions", \citet{Gaskell99} polled the 
conference attendees on the question "What drives Boroson-Green Eigenvector 1?"
The overwhelming consensus was "Don't Know", which received 68\% of the votes.

Of course, the eigenvectors are merely a mathematical construct to describe and to reduce 
the dimensionality of the object-to-object variations.  They are orthogonal by definition,
but their relationship to real physical properties may be complex or non-existent.
It is tempting, however, to use them and to explore their relationship to physical 
parameters because (a) by reducing the dimensionality they allow models to be parameterized
in simpler ways, (b) they link together diverse properties and provide more robust tests
of such models, (c) they increase the "signal-to-noise" by allowing the merging of multple
samples.  

By far the most popular interpretation has been that PC1 is highly correlated with $L/L_{Edd}$,
the Eddington ratio.  This was put forward by BG92 as the basis for a picture in which the vertical
structure of the accretion disk, governed by the Eddington ratio, drives line strengths and continuum
components through its illumination of broad-line clouds and an extended narrow-line region.  Other
factors which have been discussed as possibly playing a role in the properties that PC1 comprises are
black hole spin and orientation.  

In this paper, we explore the relation between the principal components that
describe the observed properties and determinations of the physical
properties.  We begin by reviewing the observed correlations, including
a simple visualization of the PC1 and PC2 sequences in Section 2.
In Section 3, we adopt from the literature
methods for determining the physical properties and attempt to understand PC1 and
PC2 in terms of these.  In Section 4, these relations are extended to include new
samples of radio-loud objects, and it is demonstrated that a consistent picture of the relation between
observables and physical properties emerges.  In Section 5, this picture is further 
explored as a context for the classification of QSOs, including the expected effects of orientation.  
Conclusions are summarized in Section 6.

\section{The Two Principal Components}

BG92 obtained spectra covering the region $\lambda\lambda4300-5700$ for all
87 QSOs in the BQS catalog having redshifts less than 0.5.  Measurements
of the strengths of Fe II, $H\beta$, [O III]$\lambda5007$, He II$\lambda4686$,
were combined with a four-dimensional parameterization of the H$\beta$ line
and broad-band continuum information, including $M_V$, $\alpha_{ox}$, and log
R, the ratio of radio-to-optical flux density.  Principal component
analysis performed on this dataset yielded the result that most of the
variance in the tabulated line and continuum properties was contained in
two principal components or eigenvectors, which we call PC1 and PC2.

PC1 is dominated by the inverse correlation between the strengths of Fe II
and [O III].  Objects that have strong Fe II and weak [O III] also tend to
be radio quiet, and to have $H\beta$ lines that are narrower and blue
asymmetric (more flux on short wavelength shoulder of the line).  Objects that
have weak Fe II and strong [O III] tend to be radio loud (though not exclusively), 
and to have $H\beta$
lines that are broader and red asymmetric.  Note that none of these
properties is completely correlated with PC1; although all of the extreme Fe
II-weak/[O III]-strong objects are radio-loud, for example, about half of the
radio-loud objects are mixed with radio-quiet objects in the same half of
the PC1 distribution.

PC2 is dominated by luminosity and its anticorrelation with the strength of
He II.  There is also a moderately strong relation with $\alpha_{ox}$ such that
lower luminosity objects (which have strong He II emission) rise more steeply
from their optical continuum into the x-rays.

For this study, the principal component analysis of BG92 has been repeated, using 
updated measurements of a few of the
observables.  All the $\alpha_{ox}$ values fom BG92 have been replaced with those
given by \citet{Brandt00}.  Two incorrect $H\beta$ FWHM values have
been replaced: PG 1307+085 (5320 km/s) and PG 2304+042 (6500 km/s) \citep{Laor00}.
Also, PG 1211+143 has been changed from radio-loud to radio-quiet \citep{Kellerman94}.
The variable that was called Peak $\lambda$5007, a measure of the relative
heights of the peaks of the [O III] and $H\beta$ lines, has been replaced with
a variable that is less dependent on the resolution of the observation.  The
new Peak $\lambda$5007, designated Peak2 $\lambda$5007, is given by EW
$\lambda$5007 / EW $H\beta \times H\beta$ FWHM / 1000.  This has
characteristics quite similar to Peak $\lambda$5007 (correlation coefficient r
= 0.91 between the two variables), but can be determined consistently and accurately
from observations with other instruments.

The principal component analysis was carried out using Vista, a freeware
package written and distributed by Dr. Forrest Young of the University of North
Carolina, Department of Psychology (http://www.visualstats.org).  The principal 
components that emerged from
this analysis are very similar to those found by BG92.
Figure 1 presents the distribution of the 87 low-redshift PG objects in the new
PC1-PC2 space, with different symbols used for narrow-line Seyfert 1's (NLS1s; solid circles), 
broad absorption-line QSOs (BALQSOs; solid triangles), other radio-quiet (solid squares), 
flat-spectrum radio-loud
(open triangles), and steep-spectrum radio-loud (open circles) objects.  Because the PG sample 
has a high incidence of relatively narrow line objects, a more restrictive definition of a NLS1 is
used: having H$\beta$ FWHM $<$ 1500 km $s^{-1}$ rather than the usual limit of 2000 km $s^{-1}$.

Table 1 gives information for evaluating PC1 and PC2.  Listed here are the variables
used (see BG92 for definitions), the mean and standard deviation for each variable 
in the PG sample, and the coefficients.  The PC1 and PC2 values are given by
the linear combination of the normalized variables: 
$PC1 = \sum_{}{c_{1,i}(x_i-\mu _i)/\sigma_i}$ where $c_{1,i}$ is the coefficient for the
ith variable and the first principal component, $x_i$ is the measurement of the ith
variable, and $\mu_i$ and $\sigma_i$ are the mean and standard deviation of the ith
variable.

It is interesting to note that although radio-loud objects are
distinguished clearly by neither PC1 nor PC2, almost all of them fall in the
lower right corner of the figure.  Also note that BALQSOs and NLS1 objects both
fall at the low PC1 end of the diagram, but at opposite ends of PC2. The four
corner regions marked in Figure 1 serve
to isolate the objects with extreme values of PC1 and PC2.  In order to get a
visual impression of the relationship between the principal components and the
spectra, we have averaged the spectra of the objects in each of the corner
regions.  Figure 2 shows the average spectra of the objects in the four extreme regions.
 
It is also interesting that while the increase in He II$\lambda4686$
strength is visible from bottom to top on the right hand (weak Fe II)
side, it is also clearly present in the objects that have strong Fe II
and weak [O III].  Figure 3 shows the very broad He II$\lambda4686$ line that
emerges when the spectra of the low PC1 high PC2 objects are averaged, but after
removing Fe II and continuum contributions.
 
BG92 proposed that PC1 is related to the fraction of the Eddington
luminosity at which the object is emitting.  This speculation arose
from a very qualitative picture of accretion disk structure and
radiative transfer, in which, as the accretion rate increases to a
value close to the Eddington limit, the disk puffs up vertically,
supported by radiation pressure.  The expected consequences of this include
a large x-ray heated volume that could generate the Fe II emission, and
substantial shielding of the extended narrow-line region from UV
ionizing radiation, resulting in weaker [O III] emission.  Subsequently,
some support for this explanation has been provided by studies (e.g.,
\citet{Brandt99}) of NLS1s, which
lie at one extreme of PC1.  These objects show a strong soft x-ray excess,
which is attributed to thermal emission from a viscously-heated accretion
disk that results at an accretion rate close to the Eddington limit
\citep{Pounds95}.  PC2 is presumably related predominantly
to the accretion rate itself, since it is strongly correlated with the
optical luminosity.
 
\section{Black Hole Masses}

While the masses of black holes in active galaxies have not yet been
determined through the spatially resolved spectroscopy that provides the
most convincing measurements for nearby non-active galaxies, a number
of studies have explored derivative methods.  \citet{McLeod01}
and \citet{McLeod99} use the assumption that the black hole mass is a
constant fraction of the bulge luminosity and show that sensible
fractions (a few percent to a few tens of percent) of the Eddington
luminosity result.  \citet{Kaspi00} and \citet{McLure01} use
reverberation mapping to determine the radius of the emitting material
and then use emission line widths to estimate the black hole mass.  The
samples used in these studies are small but there is some overlap with
the BG92 sample.
 
A number of studies (\citet{Kaspi96}, \citet{Laor98}, \citet{McLure01}, 
\citet{MF01}) have developed 
formalisms that relate the black hole mass to the emission line width 
and some measure of the luminosity through the assumption that the
BLR clouds' motion is virialized and using the relationship between
$R_{BLR}$ and luminosity found by \citet{Kaspi96} from reverberation 
mapping.  Although this approach no doubt represents a simplified view 
of the relationships, it has the
advantage that it provides a uniform and consistent estimate of black
hole mass from a sample that has uniform and consistent measturements
of the input parameters.   For this study, we adopt the assumptions and
parameters advocated by \citet{MF01}: 
$R_{BLR}$ = 32.9($\lambda L_{5100}/10^{44} erg s^{-1})^{0.7}$ light days 
\citep{Kaspi00} and $v_{BLR} = \sqrt{3}/2 FWHM(H\beta)$.  Using the
BG92 $H\beta$ FWHM values and the \citet{Neugebauer87} spectrophotometry
in these formulae, black hole masses can be computed for all 87 objects
in the BG92 sample.

We now would like to test whether any relationship exists between the
principal components of the PG dataset and the values of $M_{BH}$ and
$L/L_{Edd}$ derived in this way.  Because the prescription for $M_{BH}$
and $L/L_{Edd}$ use $v_{BLR}$ and $L_{5100}$, which are derived from 
(or highly correlated with)
$H\beta$ FWHM and $M_V$ respectively, we perform two additional
principal component analyses, one excluding $H\beta$ FWHM as an input
variable and one excluding both $H\beta$ FWHM and $M_V$ as input
variables.  Note that $H\beta$ FWHM but not $M_V$ figure prominently in
the first principal component, the one that we suspect is driven by
$L/L_{Edd}$.  Table 2 gives the correlation coefficients among
$L/L_{Edd}$, $M_{BH}$, and the two principal components of each of the
three analyses.  It is clear that $L/L_{Edd}$ is most highly correlated
with PC1; correlation coefficients of 0.53, 0.45, and 0.45 are found for
the three analysis.  These correspond to chance probabilities less
than 0.01\%.  Note also that $M_{BH}$ is highly correlated with both
principal components.  
 
\section{Enlargement of the sample}

The PG sample, although well defined, contains only a small number of
radio-loud QSOs.  In order to work with a sample that will allow
conclusions to be drawn about the differences between radio-quiet and
radio-loud objects, we supplement the BG92 measurements with two
radio-loud samples, 46 objects from \citet{Corbin97} and 29
additional objects from \citet{Brotherton96}.  A few of the objects 
from those samples in which the
S/N was obviously poor, or in which [O III] $\lambda$5007 could
not be accurately measured were excluded.  Measurements of the
emission-line parameters including the
strengths of $H\beta$, [O III] $\lambda$5007, and Fe II and the width,
shape, shift, and asymmetry of $H\beta$ are drawn from
those papers.  Continuum properties such as $M_V$, $\alpha_{ox}$, and
log R are drawn from those papers or references therein.  In a few
cases, we have updated these values or filled in missing values through
searches of the more recent literature.  

Having tabulated the known values for these 75 additional objects, the
approach taken toward combining the samples was not to repeat the PCA
with the entire dataset.  Because the new objects represent very
different selection criteria than the original PG sample, the variance
in the total sample would be dominated by the differences between
the original UV-excess selected objects that are predominantly radio-quiet 
and of lower luminosity and the new radio-selected objects that are at 
somewhat higher redshift and higher luminosity.  The goal of increasing 
the sample is to
use the tools that have been derived from the PG analysis to better
understand what happens when the extent of parameter space is increased.
Therefore, the projections of the new objects on the principal components 
derived in the previous section were evaluated (using the coefficients given
in Table 1) and the new PC1 vs. PC2
diagram is shown in Figure 4.  

In evaluating the new objects in terms of the original principal
components, some of the variables that are elements of the definitions
of the principal components were not measured.  Specifically, none of 
the new objects have measurements of the equivalent width of He II 
$\lambda$4686, and some do not have measurements of Fe II,
$\alpha_{ox}$, $H\beta$ shift, or $H\beta$ shape.  In these cases, a
conservative approach was used.  It was assumed that the missing value
was equal to the mean value of the PG sample.  This has the effect of
eliminating any influence that a missing variable would have on the
position of the object in PC1-PC2 space.  Another possible approach
would have been to use the established correlations between variables with
missing values and other variables in order to estimate the missing
values.  This might, however, have the effect of unrealistically exaggerating any
distinctions between the radio-loud and radio-quiet objects, and so the
more conservative approach was adopted.

Figure 4 shows a dramatic difference between the positions of radio-loud
(open symbols)
and radio-quiet (solid symbols) objects.  With almost no overlap, the two sets of
objects (including on the radio side both steep-spectrum and flat-spectrum objects)
can be divided by a straight line, drawn by eye as a dashed line
in figure 4.  

Because log R, the ratio of radio to optical flux density is a
contributor to the principal components, one might imagine that it is the
very fact that the radio-loud objects have large log R that is moving
them to the lower right part of the PC1-PC2 diagram.  To test this, the
projections of the objects on the principal components was reevaluated,
this time with no contribution from log R.  These "radio-free"
values are shown in figure 5.  As expected, the radio-loud objects have
moved up and to the left, but not nearly far enough to eliminate the
distinction between radio-loud and radio-quiet.  The same dashed line as
in figure 4 is shown in figure 5 for reference.

What physical parameter causes this separation?  By using $H\beta$ FWHM
and optical luminosity to compute $M_{BH}$ for the PG sample, \citet{Laor00} 
finds that radio-loud QSOs are associated with the most massive
black holes.  This result has been confirmed and refined by \citet{Lacy01}
who added a radio-selected sample to the PG objects.  \citet{Lacy01}
find that radio luminosity scales proportional to
$M_{BH}^{1.9}(L/L_{Edd})^{1.0}$.  

Using the same approach as \citet{Laor00}, the black hole masses for this sample
can be evaluated.  These are shown by the color coding of the points in the PC1-PC2
plane in figure 6.  The objects have been binned in ranges of 0.5 in log $M_{BH}$ and
the average positions of all objects in each of those bins are shown as the large
colored plus signs in figure 6.  It is clear that the trend in $M_{BH}$ parallels the trend 
in radio loudness.  This is exactly the expectation, of course, if PC1 is driven primarily 
by $L/L_{Edd}$ (proportional to L/M) and PC2 is driven primarily by accretion rate 
(proportional to L).

In addition to the trend of increasing black hole mass from upper left to lower right, another
striking feature of figure 6 is that the transition between radio-quiet and radio-loud objects
is much more sharply defined than are the black hole mass bins.  Thus, both the cyan and 
magenta objects (9.0 $<$ log $M_{BH} <$ 10.0) appear to span the dividing line between
radio-quiet and radio-loud.  One possible explanation for this is that the PC1-PC2 plane 
provides a superior method for determining black hole mass than does the simple
formula that uses only H$\beta$ FWHM and luminosity.  Another possibility, though, is that
there is a correlation between the radio emission itself and one or more of the other 
properties that consitute the principal components.  For example, it is possible that radio jets 
trigger star formation which produces an enhanced level of [O III] emission.  A test for this
would be to compare the [O III] emission strength for radio-loud and radio-quiet objects with
the same Fe II emission strength.   Such a test shows no difference in [O III] between the two
groups, however.

\section{Discussion: A Paradigm for AGN classification}

Figure 6 suggests a relatively simple diagram, shown in figure 7, that links together physical properties 
and observables.  This diagram, which provides a physical basis for classification of different types of AGNs,
something like an HR diagram does, plots luminosity or accretion rate against Eddington ratio.  The observational
analogs of these properties are PC2 and PC1.  Note that the theoretical quantities are not orthogonal but are
related linearly, although PC2 and PC1 are orthogonal mathematically within the PG sample.  Lines of
constant black hole mass are diagonal lines in this diagram.  

The extreme regions along both axes may be restricted because of either physical constraints (e.g., difficulty 
in accreting above the Eddington limit) or observational constraints (e.g., too low a luminosity will preclude 
classification as an AGN).  Within the allowed region it can be seen that radio loud QSOs will be found 
predominantly in elliptical galaxies, systems hosting black holes of greater mass,  while radio-quiet QSOs 
will be found in lower mass spheroidal systems -- i.e., generally spirals -- as they have lower mass black holes. The
most extreme low black hole mass objects are the NLS1s, falling at the upper left of the PC1-PC2 diagram.

BALQSOs apparently fall at the lower left corner, that is, high Eddington ratio and high luminosity.  Since this
position is near the dividing line between radio-loud and radio-quiet, one could conjecture that radio-loud BALQSOs 
would have extremely high luminosity, causing them to fall below that dividing line.  This is consistent
with the most promising candidates for radio-loud BALQSOs, such as FIRST J155633.8+351758 \citep{Najita00}.
Thus, the scarcity of radio-loud BAL QSOs would be due to the relatively small region of parameter space 
that is available when an object is to be a BAL (high Eddington ratio, high accretion rate) and radio-loud (large
black hole mass).

As noted by \citet{Laor00},  it might be expected that three parameters, black hole mass, Eddington ratio, and
orientation, would determine the observable characteristics of AGNs.  While the picture developed in this 
paper includes the first two, there has been little indication of anything that could be interpreted as orientation.
This is, in some sense, in conflict with the "standard unification model" in which orientation plays a major role
in providing context for the classification of AGNs \citep{Antonucci93,Urry95}.  The unification model, based on the ideas of relativistically
beamed emission and a thick (vertically and optically) torus, receives support from scattered broad line emission
seen in some Seyfert 2 galaxies, the "alignment effect" seen in high redshift radio galaxies, as well as a number 
of consistency arguments applied to various statistical samples.  

Within the samples considered in this study, the most obvious groups to explore for orientation
effects are the core-dominated (flat-spectrum) and lobe-dominated (steep-spectrum) radio-loud QSOs.  
It has been conjectured that if a radio-loud QSO is seen along its jet, it will 
appear as a core-dominated source, with the core radio emission enhanced by relativistic beaming.  If the jet is 
close to the plane of the sky, however, the lobes will dominate, and will be well apart from the cental source.  
Studies comparing the two types suffer from the difficulty of picking samples that include objects of similar
intrinsic properties.  

Although Figure 6 shows little obvious separation between the core-dominated (open triangles) and 
lobe-dominated (open circles) objects, a Kolmogrov-Smirnov test indicates a significant difference 
(chance probability $<$ 0.5\%) in the distribution of PC1, with core-dominated
objects having lower values of PC1 by about 1.0 in the median than lobe-dominated objects.  Of course, log R is
a component of PC1 (as log R increases, PC1 increases), but the median log R value for the core-dominated objects 
is larger than that of the lobe-dominated objects: 3.14 as opposed to 2.56
No significant difference is seen in PC2.  

What would be expected? If core-dominated objects have weaker intrinsic radio luminosity but their emission is
enhanced by beaming and if the orientation has no effect on any property other than log R used in our analysis,
then this is qualitatively a consistent finding.  Within this simplistic picture, an enhancement of 2.8 in log R is 
required to explain the location of the core-dominated objects.  This is extremely speculative as it depends on
(a) unquantifiable selection effects in the samples considered here, (b) unknown additional effects that orientation
might have on other properties that went into the analysis, and (c) a derivation of principal components that was
based on the assumption that the log R values of core- and lobe-dominated objects have the same intrinsic meaning.

\section{Conclusions}

A revised statistical analysis of the properties of the low-redshift PG QSOs has been performed, with updated
measurements.  Two principal components have been identified, corresponding, as in BG92, to the Fe II--[O III]
anticorrelation and to the luminosity.  Black hole masses for this sample have been estimated using
the formula derived by \citet{Laor00}.  These indicate that the two principal components correspond closely to the 
Eddington ratio and the accretion rate, respectively.  Two additional radio-loud samples have been added to
the analysis, by using the principal component definitions derived from the PG sample.  Inspection of the PC1-PC2
diagram for this enlarged sample shows that:
\begin{enumerate}
\item Diagonal lines in the PC1-PC2 diagram represent lines of constant black hole mass.

\item Radio-loud and radio-quiet objects are well separated by such a line, indicating that radio-loudness
 is determined by black hole mass.

\item NLS1s lie at the corner of the diagram indicative of having the lowest black hole masses.

\item BAL QSOs lie in the high Eddington ratio, high accretion rate corner of the diagram, suggesting that the
rare radio-loud BAL QSOs must have extremely high accretion rates.

\item Comparison of the location of core-dominated and lobe-dominated radio-loud QSOs suggests that orientation
manifests itself as a shift in PC1, perhaps due primarily to relativistic beaming of radio emission.  With this assumption,
an enhancement factor of 2.8 in the log is derived.
\end{enumerate}

\acknowledgments

I wish to thank Richard Green and Michael Brotherton for helpful discussions,
and Ari Laor for the use of his database.  This research has made use of the
NASA/IPAC Extragalactic Database (NED) which is operated by the Jet Propulsion
Laboratory, California Institute of Technology, under contract with the
National Aeronautics and Space Adminsitration.

\clearpage

\begin{figure}
\figurenum{1}
\plotone{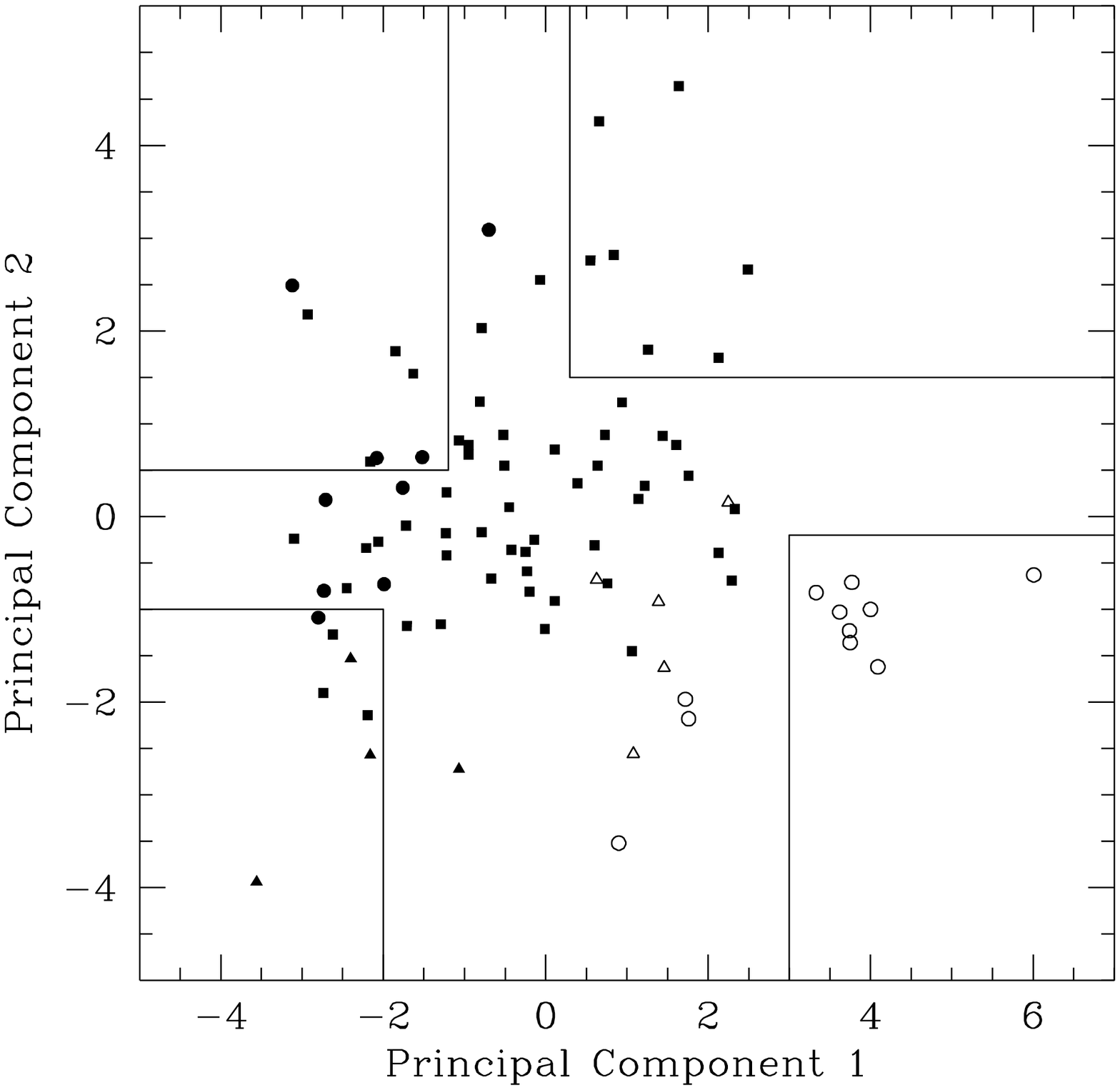}
\caption{Distribution of the 87 low-redshift PG objects with respect to the two
principal components.  Symbols are solid triangles: BALQSOs, solid circles:
NLS1s, solid squares: other radio quiet QSOs, open triangles: flat-spectrum
radio-loud QSOs, and open circles: steep-spectrum radio-loud QSOs. The four
marked off regions identify the objects whose spectra have been averaged to
show the characteristics of the extremes of the two principal components.}
\end{figure}

\begin{figure}
\figurenum{2}
\plotone{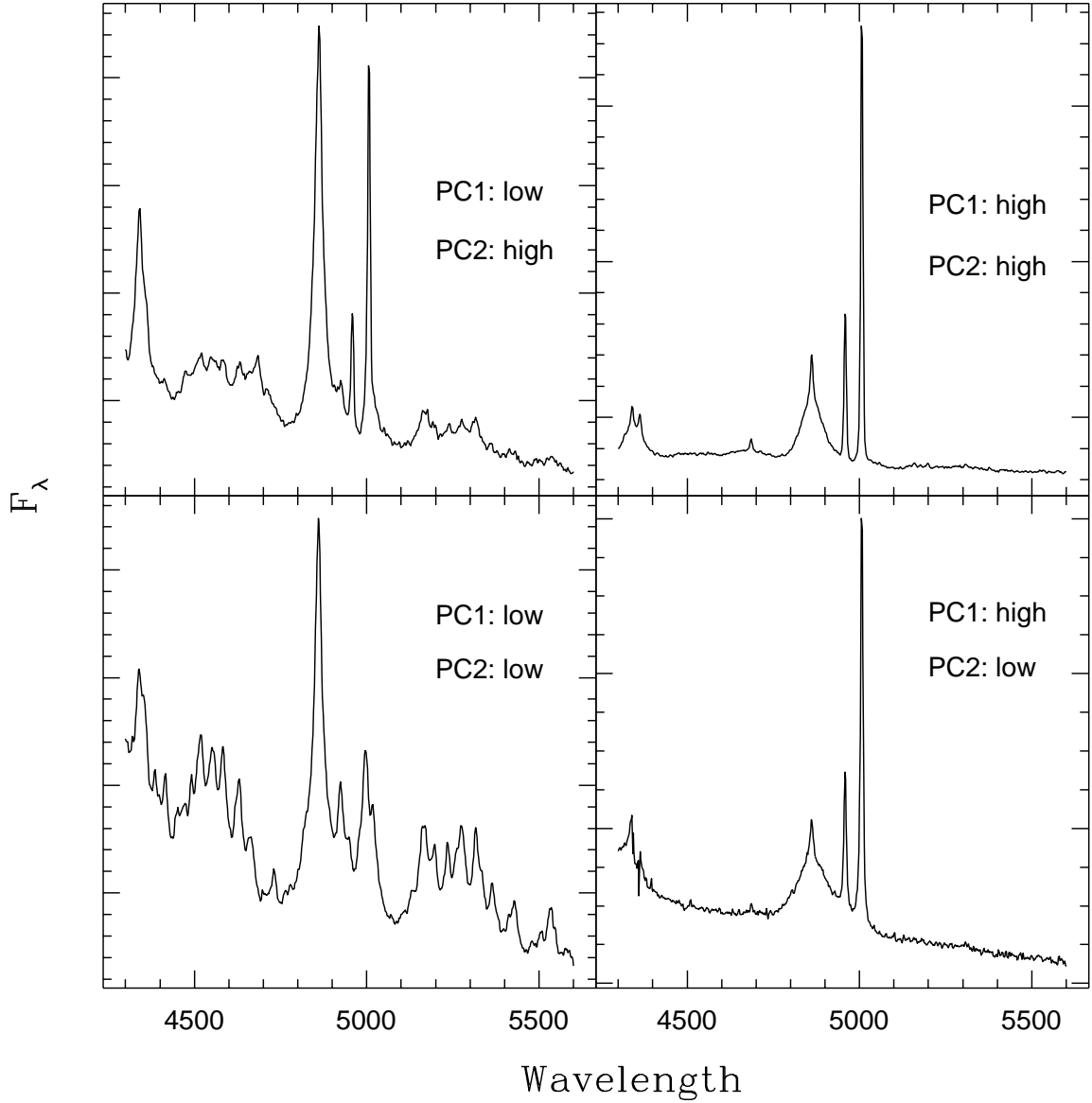}
\caption{Average spectra of the PG objects having extreme values of the two
prinicpal components. Each quadrant shows the average spectrum of the objects
in the corresponding corner of figure 1.}
\end{figure}

\begin{figure}
\figurenum{3}
\plotone{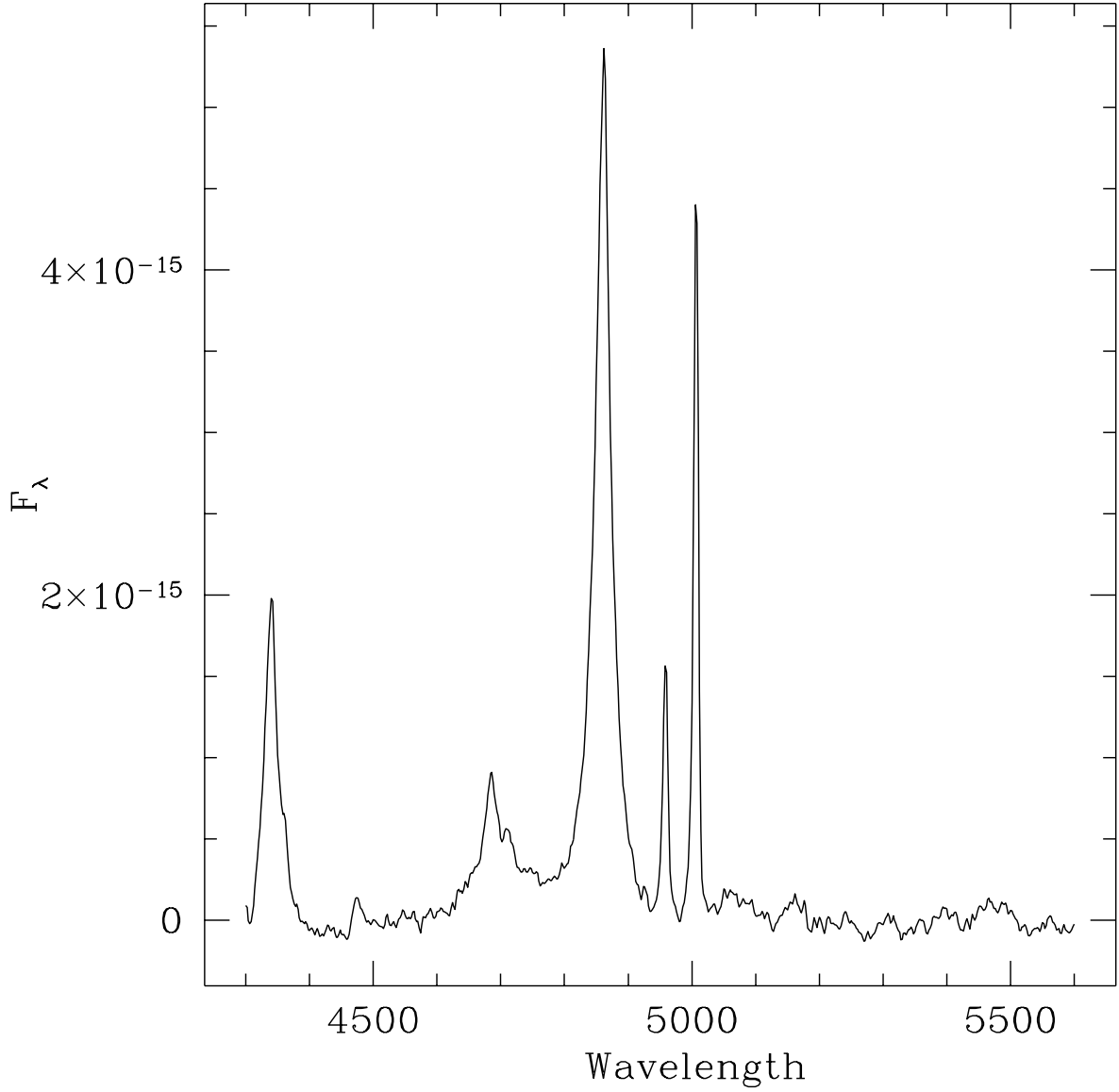}
\caption{Average spectrum of the objects in the upper left (low PC1, high PC2) 
corner of figure 1, but with Fe II and continuum subtracted first.  Note the
strong, broad He II $\lambda4686$ emission line previously masked by the Fe II
emission.}
\end{figure}

\begin{figure}
\figurenum{4}
\plotone{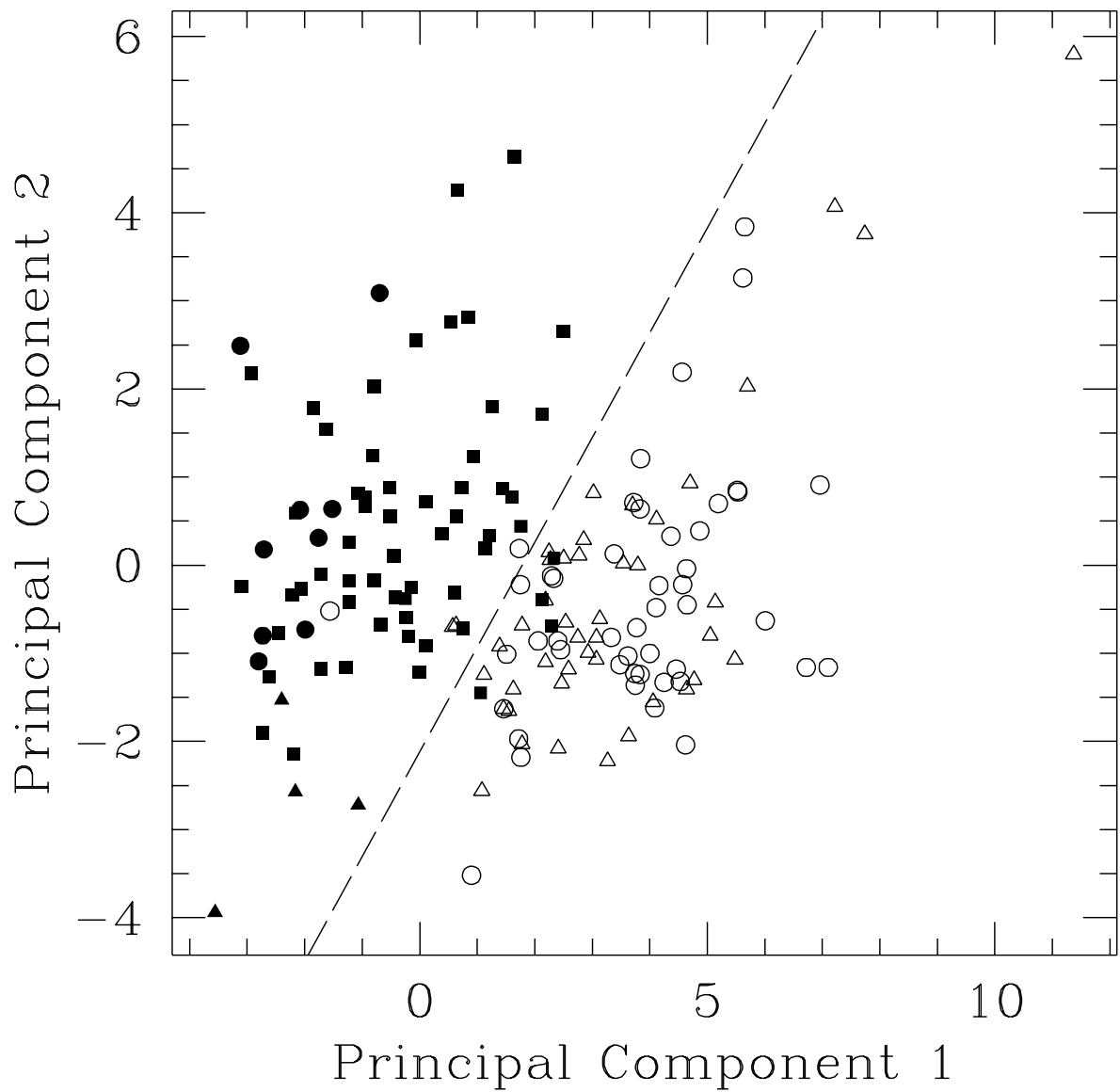}
\caption{Distribution of the enlarged sample of 162 objects with respect 
to the two
principal components.  Symbols are solid triangles: BALQSOs, solid circles:
NLS1s, solid squares: other radio quiet QSOs, open triangles: flat-spectrum
radio-loud QSOs, and open circles: steep-spectrum radio-loud QSOs. The
dashed line (drawn by eye) separates radio-loud and radio-quiet
objects.}
\end{figure}

\begin{figure}
\figurenum{5}
\plotone{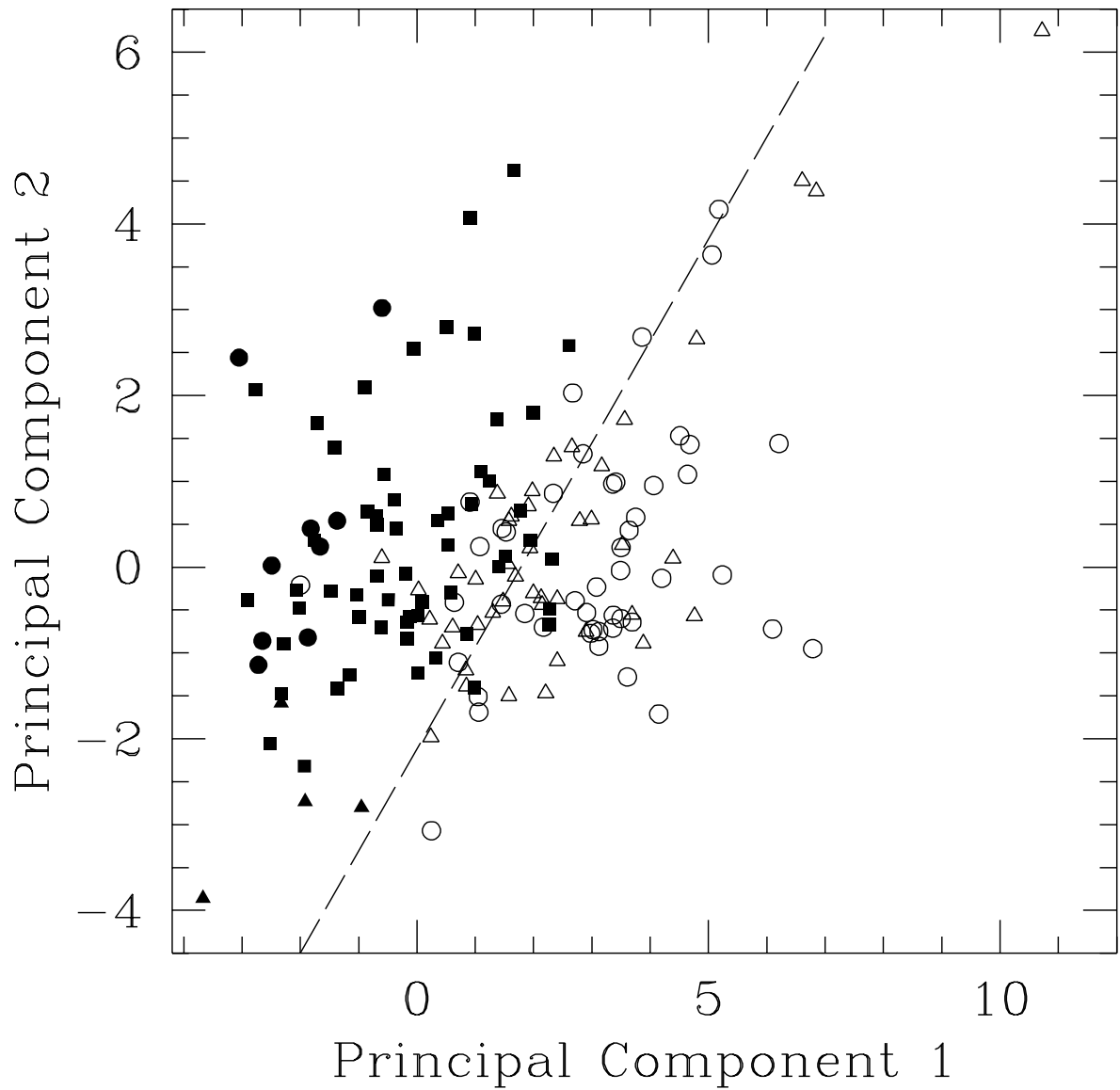}
\caption{Same as figure 4 but with the effect of Log R removed from the
principal components.  Symbols are solid triangles: BALQSOs, solid circles:
NLS1s, solid squares: other radio quiet QSOs, open triangles: flat-spectrum
radio-loud QSOs, and open circles: steep-spectrum radio-loud QSOs. The
dashed line is the same line as in figure 4, showing that the radio-loud
objects have moved up and to the left, but are still well separated from
the radio-quiet objects.}
\end{figure}

\begin{figure}
\figurenum{6}
\plotone{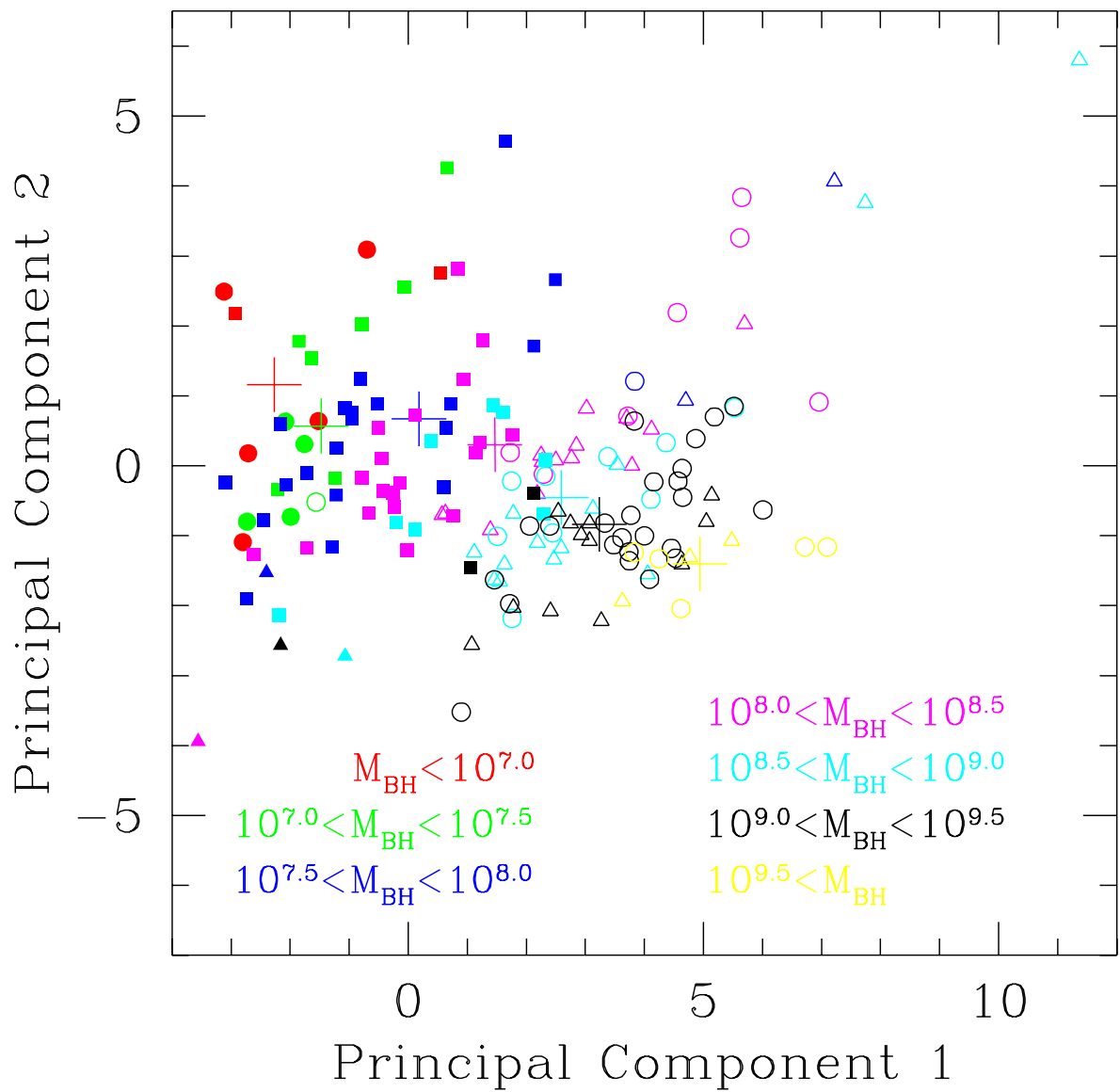}
\caption{Distribution of objects in the PC1-PC2 plane, now color coded according
to black hole mass as derived from the \citet{MF01} formula.  Symbols are
as in figures 4 and 5.  Range of black hole mass denoted by each color is given
in legend at bottom.   Large plus signs show average position of objects in each
black hole mass bin. }
\end{figure}

\begin{figure}
\figurenum{7}
\plotone{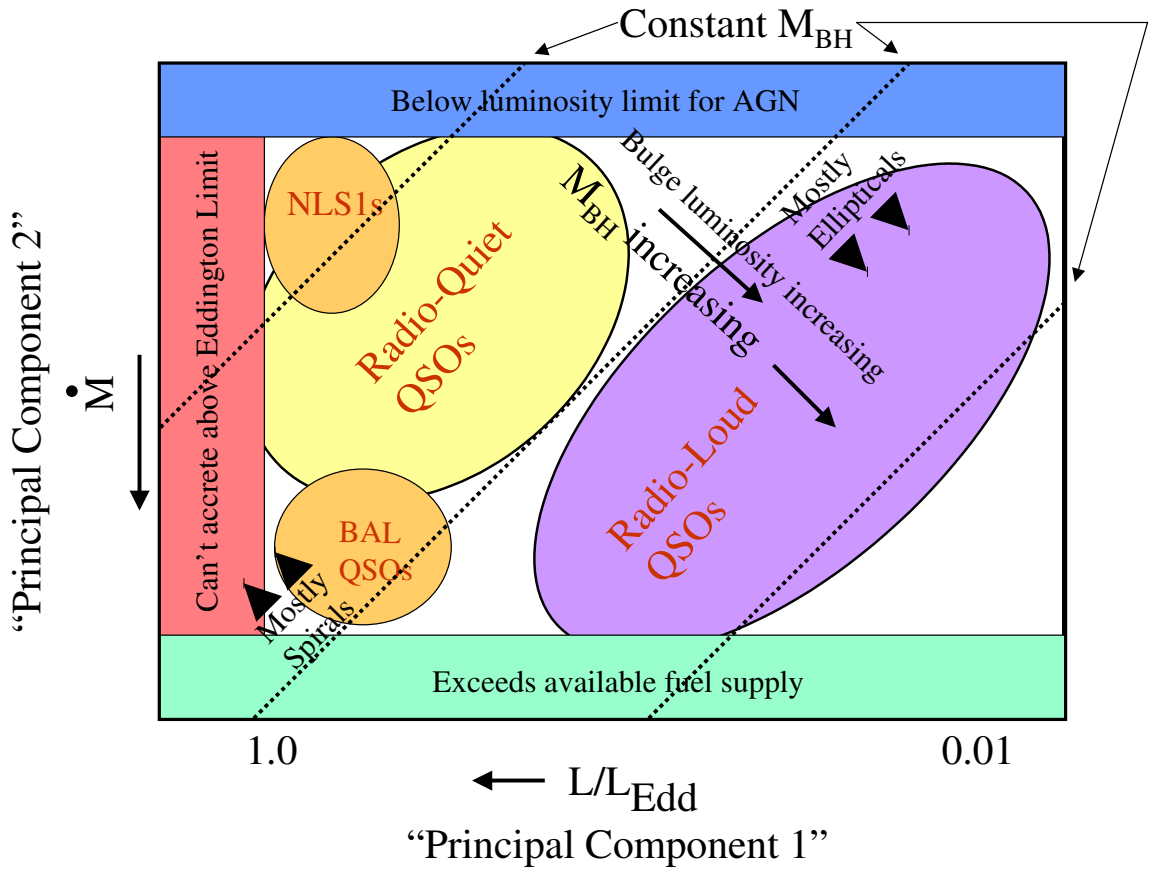}
\caption{Interpretive diagram showing how PC1-PC2 plane provides basis for classification
of AGNs. }
\end{figure}





\clearpage

\begin{deluxetable}{lrrrr}
\tablecaption{Coefficients for PC1 and PC2}
\tablehead{
\colhead{Variable} & 
\colhead{Mean}   & 
\colhead{Standard Deviation}   &
\colhead{Coefficient for PC1} &
\colhead{Coefficient for PC2} 
}
\startdata

$M_V$&$-$24.16&1.53&$-$0.1777&$+$0.5389\\
EW H$\beta$ &95.55&37.31&$+$0.1046&$+$0.0807\\
R $\lambda$ 5007&0.26&0.22&$+$0.2645&$+$0.3446\\
R $\lambda$4686&0.10&0.10&$-$0.0765&$+$0.4910\\
R Fe II&0.55&0.38&$-$0.4001&$-$0.1164\\
H$\beta$ FWHM&3777.00&1994.00&$+$0.3719&$-$0.1039\\
$\alpha_{ox}$&1.58&0.23&$-$0.2097&$-$0.2919\\
H$\beta$ shift&0.00&0.07&$+$0.0083&$-$0.0987\\
H$\beta$ shape&1.16&0.09&$-$0.0774&$-$0.2887\\
H$\beta$ asymm&$-$0.01&0.10&$-$0.3311&$-$0.1004\\
$M_{[O III]}$&$-$27.02&1.62&$-$0.4137&$+$0.2030\\
Peak2 $\lambda$5007&1.04&1.17&$+$0.3760&$+$0.1887\\
Log R&$-$0.04&1.23&$+$0.3333&$-$0.2317\\

\enddata


\end{deluxetable}

\clearpage

\begin{deluxetable}{lrrrrrrrr}
\tablecaption{Linear Correlation Coefficients between Principal
Components, $M_{BH}$ and $L/L_{Edd}$\label{tbl-2}}
\tablehead{
\colhead{    } & 
\colhead{$L/L_{Edd}$}   & 
\colhead{log $M_{BH}$}   &
\colhead{PC1\tablenotemark{a}} &
\colhead{PC2\tablenotemark{a}} & 
\colhead{PC1\tablenotemark{b}} & 
\colhead{PC2\tablenotemark{b}} & 
\colhead{PC1\tablenotemark{c}} & 
\colhead{PC2\tablenotemark{c}} 
}
\startdata

$L/L_{Edd}$&1.00 &$-$0.56 &0.53 &$-$0.04 &0.45 &0.04 &0.45 &$-$0.07 \\
log $M_{BH}$&-0.56 &1.00 &$-$0.67 &$-$0.57 &$-$0.56 &0.60 &$-$0.47 &$-$0.48 \\
PC1\tablenotemark{a}&0.53 &$-$0.67 &1.00 &0.00 &0.99 &$-$0.08 &0.92 &0.17 \\
PC2\tablenotemark{a}&$-$0.04 &$-$0.57 &0.00 &1.00 &$-$0.09 &$-$0.99 &$-$0.22 &0.88 \\
PC1\tablenotemark{b}&0.45 &$-$0.56 &0.99 &$-$0.09 &1.00 &0.00 &0.99 &0.11 \\
PC2\tablenotemark{b}&0.04 &0.60 &$-$0.08 &$-$0.99 &0.00 &1.00 &0.13 &$-$0.91 \\
PC1\tablenotemark{c}&0.45 &$-$0.47 &0.97 &$-$0.22 &0.99 &0.13 &1.00 &0.00 \\
PC2\tablenotemark{c}&$-$0.07 &$-$0.48 &0.17 &0.88 &0.11 &$-$0.91 &0.00 &1.00 \\

\enddata


\tablenotetext{a}{All variables included}
\tablenotetext{b}{H$\beta$ FWHM excluded}
\tablenotetext{c}{H$\beta$ FWHM and $M_V$ excluded}

\end{deluxetable}




\begin{thebibliography}{}
\bibitem[Antonucci(1993)]{Antonucci93} Antonucci, R. 1993, \araa, 31, 473.
\bibitem[Boroson and Green(1992)]{BG92} Boroson, T.A. and Green, R.F. 1992,
    \apjs, 80, 109
\bibitem[Brandt and Boller(1999)]{Brandt99} Brandt, W.N. and Boller, T. 1999,
    in ASP Conf. Ser. 175, Structure and Kinematics of Quasar Broad Line Regions, 
    ed. C. M. Gaskell, W.N. Brandt, M. Dietrich, D. Dultzin-Hacyan, and M. 
    Eracleous (San Francisco: ASP), 265
\bibitem[Brandt, Laor, and Wills(2000)]{Brandt00} Brandt, W.N., Laor, A., and
    Wills, B.J. 2000, \apj, 528, 637
\bibitem[Brotherton(1996)]{Brotherton96} Brotherton, M.S. 1996, \apjs, 102, 1
\bibitem[Corbin(1997)]{Corbin97} Corbin, M.R. 1997, \apjs, 113, 245
\bibitem[Ferrarese and Merritt(2000)]{Ferrarese00} Ferrarese, L., and Merritt,
    D. 2000, \apjl, 539, L9
\bibitem[Gaskell(1999)]{Gaskell99} Gaskell, C. M. 1999, in ASP Conf. Ser. 175, 
    Structure and Kinematics of Quasar Broad Line Regions, 
    ed. C. M. Gaskell, W.N. Brandt, M. Dietrich, D. Dultzin-Hacyan, and M. 
    Eracleous (San Francisco: ASP), 423
\bibitem[Gebhardt et al.(2000)]{Gebhardt00} Gebhardt, K., Bender, R., Bower,
    G., Dressler, A., Faber, S.M., Filippenko, A.V., Green, R., Grillmair, C., Ho,
    L.C., Kormendy, J., Lauer, T.R., Magorrian, J., Pinkney, J., Richstone, D., and
    Tremain, S. 2000, \apjl, 539, L13
\bibitem[Kaspi et al.(1996)]{Kaspi96} Kaspi, S., Smith, P.S., Maoz, D., Netzer,
    H., and Jannuzi, B.T. 1996, \apjl, 471, L75
\bibitem[Kaspi et al.(2000)]{Kaspi00} Kaspi, S., Smith, P.S., Netzer, H., Maoz,
    D., Jannuzi, B.T., and Giveon, U. 2000, \apj, 533, 631
\bibitem[Kellerman et al.(1994)]{Kellerman94} Kellerman, K.I., Sramek, R.A.,
    Schmidt, M., Green, R.F., and Shaffer, D.B. 1994, \aj, 108, 1163
\bibitem[Lacy et al.(2001)]{Lacy01} Lacy, M., Laurent-Muehleisen, S.A.,
    Ridgway, S.E., Becker, R.H., and White, R.L. 2001, \apjl, 551, L17
\bibitem[Laor(1998)]{Laor98} Laor, A. 1998, \apjl, 505, L83
\bibitem[Laor(2000)]{Laor00} Laor, A. 2000, \apjl, 543, L111
\bibitem[Mcleod and McLeod(2001)]{McLeod01} McLeod, K.K. and McLeod, B.A. 2001,
    \apj, 546, 782
\bibitem[McLeod, Rieke, and Storrie-Lombardi(1999)]{McLeod99} Mcleod, K.K.,
    Rieke, G.H., and Storrie-Lombardi, L.J. 1999, \apjl. 511, L67
\bibitem[McLure and Dunlop(2001)]{McLure01} McLure, R.J. and Dunlop, J.S. 2001,
    \mnras, submitted (astro-ph/0009406)
\bibitem[Merritt and Ferrarese(2001)]{MF01} Merritt, D. and Ferrarese, L. 2001, in ASP Conf.
    Ser. xxx, The Central kpc of Starbursts and AGN, ed., J.H. Knapen, J.E. Beckman, I. Shlosman,
    and T.J. Mahoney (San Francisco: ASP), in press.
\bibitem[Najita, Dey, and Brotherton(2000)]{Najita00} Najita, J., Dey, A., Brotherton, M.
    2000, \aj, 120, 2859
\bibitem[Neugebauer et al.(1987)]{Neugebauer87} Neugebauer, G., Green, R.F.,
    Matthews, K., Schmidt, M., Soifer. B.T., and Bennett, J. 1987, \apjs, 63, 615
\bibitem[Pounds, Done, and Osborne(1995)]{Pounds95} Pounds, K.A., Done, C., and
    Osborne, J.P. 1995, \mnras, 277, L5
\bibitem[Sulentic, Marziani, and Dultzin-Hacyan(2000)]{Sulrev00} Sulentic,
    J.W., Marziani, P., and Dultzin-Hacyan, D., \araa, 38 521
\bibitem[Sulentic et al.(2000)]{Sulentic00} Sulentic, J.W., Zwitter, T.,
    Marziani, P., Dultzin-Hacyan, D. 2000, \apjl, 536, L5
\bibitem[Urry and Padovani(1995)]{Urry95} Urry, C.M. and Padovani, P. 1995, 
    \pasp, 107, 803

\end{thebibliography}
\end{document}